\documentclass{article}

\usepackage{arxiv}

\usepackage[utf8]{inputenc} 
\usepackage[T1]{fontenc}    
\usepackage{hyperref}       
\usepackage{url}            
\usepackage{booktabs}       
\usepackage{amsfonts}       
\usepackage{nicefrac}       
\usepackage{microtype}      
\usepackage{lipsum}		
\usepackage{graphicx}
\usepackage{natbib}
\usepackage{doi}

\usepackage{tikz}
\usetikzlibrary{fit}
\usepackage{pgfplots}
\pgfplotsset{compat=newest}
\usetikzlibrary{pgfplots.statistics}
\usetikzlibrary{positioning,shapes,arrows,shadows,patterns}

\usepackage{dcolumn}
\usepackage{multirow}
\usepackage{amssymb}

\usepackage{amsmath}
\usepackage{algorithm}
\usepackage{enumitem}
\usepackage{algpseudocode}
\algtext*{EndIf}
\algtext*{EndFunction}
\algtext*{EndFor}

\newcommand{\fbf}{\textbf}
\usepackage{xspace}
\newcommand{\etal}{{et al.\@\xspace}}
\newcommand{\fsub}[1]{\textsubscript{#1}}
\newcommand{\fsl}{\textsl}
\newcommand{\fsup}[1]{\textsuperscript{#1}}

\newcommand{\shortcite}[1]{[\citeyear{#1}]}

\newcommand{\goR}{\rightharpoondown}
\newcommand{\goL}{\leftharpoonup}
\newcommand{\goU}{\upharpoonright}
\newcommand{\goD}{\downharpoonleft}
\newcommand{\gono}{\oslash}
\newcommand{\go}{\star}

\newcommand{\citepos}[1]{\citeauthor{#1}'s [\citeyear{#1}]}

\definecolor{revision}{rgb}{0.0, 0.0, 0.0}
\newcommand{\revised}[1]{\textcolor{revision}{#1}}
\newcommand{\revisedst}[1]{}

\definecolor{revision2}{rgb}{0.0, 0.0, 0.9}
\newcommand{\revisedb}[1]{\textcolor{black}{#1}}

\newcommand{\revisedbst}[1]{}

\newcommand{\framework}{{Cha}\xspace}

\newcommand{\frameworkNeu}{{Cha Neutral}\xspace}

\usepackage{siunitx}
\usepackage{diagbox}

\title{Prosocial Norm Emergence in Multiagent Systems}

\date{} 					

\author{ {\hspace{1mm}Mehdi ~Mashayekhi} \\
	North Carolina State University\\
	Raleigh, NC 27695 \\
	\texttt{mmashay2@ncsu.edu} \\
	\And
	{\hspace{1mm}Nirav Ajmeri} \\
	University of Bristol\\
	Bristol, United Kingdom BS8 1UB \\
	\texttt{nirav.ajmeri@bristol.ac.uk} \\
	\And
	{\hspace{1mm}George  F.~List} \\
	North Carolina State University\\
	Raleigh, NC 27695 \\
	\texttt{gflist@ncsu.edu  } \\
	\And
	{\hspace{1mm}Munindar  P.~Singh} \\
	North Carolina State University\\
	Raleigh, NC 27695 \\
	\texttt{mpsingh@ncsu.edu} \\
}

\renewcommand{\headeright}{}
\renewcommand{\undertitle}{}

\begin{document}
\maketitle

\begin{abstract}

Multiagent systems provide a basis for developing systems of autonomous entities and thus find application in a variety of domains. \revisedb{We consider a setting where not only the member agents are adaptive but also the multiagent system viewed as an entity in its own right is adaptive.} 
\revisedbst{We consider a setting where not only the member agents are adaptive but also the multiagent system itself is adaptive.}
Specifically, the social structure of a multiagent system can be reflected in the social norms among its members. 
It is well recognized that the norms that arise in society are not always beneficial to its members. We focus on prosocial norms, which help achieve positive outcomes for society and often provide guidance to agents to act in a manner that takes into account the welfare of others.

Specifically, we propose Cha, a framework for the emergence of prosocial norms. 
Unlike previous norm emergence approaches, Cha supports continual change to a system (agents may enter and leave) and dynamism (norms may change when the environment changes). 
Importantly, Cha agents incorporate prosocial decision making based on inequity aversion theory, reflecting an intuition of guilt arising from being antisocial. In this manner, Cha brings together two important themes in prosociality: decision making by individuals and fairness of system-level outcomes. 
We demonstrate via simulation that Cha can improve aggregate societal gains and fairness of outcomes.

\end{abstract}

\pagenumbering{arabic}
\thispagestyle{plain}
\pagestyle{plain}


\section{Introduction}
\label{sec:introduction}

Major practical applications of information technology can be understood as multiagent systems viewed from a sociotechnical perspective in terms of a social tier (comprising social entities such as people and organizations) and a technical tier (comprising computational and other resources) \citep{IS-16:Revani}. The agents represent the social entities computationally and reflect the interests of the entities they represent. As \citet{TIST-13-Governance} points out, the governance of such systems is naturally characterized by the social norms that apply at the social tier. Indeed, norms help regulate interactions of autonomous agents \citep{Verhagen+18:NorMAS}. Accordingly, it is not surprising that norms apply in resource-sharing settings broadly \citep{Dagstuhl-NorMAS-uses-13}, such as social media \cite{Sen2018} and for road sharing by autonomous vehicles \cite{DellAnna-AAMAS19-Runtime}. In general, discussions of norms can be framed in terms of agent behavior; here, we focus on interactions \revisedb{between agents}, to clarify that we are interested in agent behaviors that are of concern to others and are visible to others.

\revisedb{
The recent burgeoning interest in AI ethics and safety has drawn increasing attention to regulating how intelligent agents act and interact \cite{Winikoff+Sardelic-21:human-rights}. In this regard, it is helpful to distinguish macro ethics (focused on a system) from micro ethics (focused on an individual agent) \citep{AIES-18:ethics}. Micro ethics is concerned with how agents interact in light of the norms \cite{Chaput+21:ethical,Hilpinen-81}, and macro ethics is concerned with what norms regulate a system, including considerations of fairness \citep{AAMAS-20:Blue-Sky,Woodgate+Ajmeri-AAMAS22-BlueSky}.
}

At its heart, a social norm defines sound or ``normal'' interactions among members of a social group, reflecting their mutual expectations. \revised{A norm can be violated by a member though at the cost of suffering either a moral or another sanction \citep{Nardin+16:Sanctioning}.} A norm can be descriptive, merely describing a social group, or prescriptive, describing how the social group \revised{should function}. Moreover, some norms can be constitutive, meaning they provide definitions of what counts as what---in essence, the rules of the game. Or, they can be regulative, meaning they constrain the interactions of the participants \citep{Boella+van-der-Torre-04}. A lot of the work on multiagent systems approaches them from the standpoint of explicit engineering, where stakeholder requirements are mapped to, among other elements, a set of norms \citep{TOSEM-20:Desen}. 


This paper adopts an equally well-established alternative perspective in which the norms are not designed \revisedb{into a system} but emerge through the interactions of the agents \citep{Hollander+Wu-SASO-11}. Specifically, this paper proposes \framework, a framework for norm emergence. Our motivations behind \framework are as follows. First, we would like to enable agents who incorporate prosocial decision making to achieve norms that avoid conflict and improve fairness. Second, we would like to tackle the challenges of flexibility in that it should be possible for the emerged norms in a multiagent system to change dynamically. Third, the process of emergence should not have to rely upon a central authority that somehow forces the agents to adopt or respect norms; instead, the norms should arise---and, when appropriate, change---in a decentralized manner solely through the interactions among the member agents. 


Next, we abstract out some key desiderata from the above motivations and discuss how \framework addresses these desiderata.

\begin{description}

\item [Prosocial.]
A prosocial attitude is an attitude to benefit another. Prosocial behavior has been well studied in sociology and social psychology \cite{Simpson+AR15:altruism,Dovidio+17:prosocial}. A prosocial behavior is when an agent performs an action that benefits others even if suboptimal to itself \cite{Paiva2018EngineeringPW,Santos2019EvolutionOC,Serramia2018}. Whereas existing approaches for norm emergence focus solely on decision making by agents, we relate social norms both to prosocial decision making and to societal outcomes such as fairness. We show how to incorporate prosociality into norm emergence to achieve norms that promote fair outcomes and improve social welfare for the members of a multiagent system.

\item [Flexible and Enduring.]
The flexibility and endurance (in other words, the longevity) of an interaction are important aspects of openness \cite{Verhagen-2001-simulation}. The membership of a multiagent system can change yet
there is continuity in learning norms. An example is Wikipedia, where
the users are autonomous and changing but can and do build on each other's work. In contrast, existing studies of norm emergence \cite{Airiau2014,sugawara-AAMAS14-emergence,Sen_Onkur2009,Yu-AAMAS13-Emergence} apply social
learning  and assume that
agents repeatedly interact in a closed system. Specifically, these approaches assume that a fixed graph is given and the agents interact with their neighbors in that graph.

\item [Dynamic.]
Dynamism refers to the idea that norms may change with changes to the environment in which the system operates \cite{Verhagen-2001-simulation,Savarimuthu-2011-Norm+Survey,Huang2016}. Existing works \cite{Mihaylov2014,Morales2018} address dynamism inadequately. For example, once norms have emerged, they remain fixed. Existing approaches, such as \citet{DellAnna-AAMAS19-Runtime,DellAnna-JAAMAS20-Runtime}, support norm change at runtime but via a centralized sanction revision, which is prone to a single point of failure. 

\item [Decentralized.]
Decentralization refers to the absence of central authority \cite{Morris+19:norm-emergence}. In a decentralized system, norms arise solely through agent interactions \cite{Mihaylov2014,Campos+13:robust}. Most existing norm emergence approaches, such as \citet{Airiau2014,Morales:2015}, involve a central authority that
determines the norms; some existing approaches, such as \citet{IJCAI-16:Silk}, use a 
hybrid scheme. \revisedbst{However, central and hybrid schemes are
vulnerable to the failure of the central portion.} \revisedb{A criticism of hybrid, and especially of centralized, systems is that they assume that an all-knowing authority is present in the system \cite{Savarimuthu-2011-Norm+Survey}. Whereas a central authority may recommend norms to achieve a societal criterion, such an approach cannot guarantee the autonomy of the participating agents. In settings with unreliable or delayed communication or where components can fail, a centralized system is fragile because it has a single point of failure.}

\end{description}

\subsection{Contributions}
\framework is a general dynamic and flexible framework for norm emergence that
promotes prosociality while supporting decentralization. We frame our contributions as an investigation of the following themes. 

\begin{description}
\item [Efficient resolution of conflicts] captures the idea that norms emerge to avoid 
conflicts that arise between the members of a multiagent system. \revised{In addition, efficiency indicates that the norms that emerge avoid conflict without damaging performance.} In other words, the norms should address the liveness-safety tradeoff \cite{AAAI-17:Kont} without, for example, achieving high safety at the cost of liveness. 

\item [Dynamic adaptation] captures the idea that the norms in a multiagent system must be sensitive to its operating environment. That is, both the conflicts and the performance opportunities that agents face depend upon the environment in which they interact with each other. 
\revisedb{In other words, in addition to the individual member agents potentially adapting to the environment, the multiagent system as a whole would adapt through new norms.}
\revisedbst{Therefore, it is important that the multiagent system be able to adapt continually to its environment in addition to the individual member agents potentially adapting to the environment.}

\item [Fairness] refers to the idea that disparities in the outcomes across the various member agents are low. In general, there is a well-known tradeoff between the aggregate outcomes for the members of a society and the fairness of the individual outcomes received by its members \cite{Adler-19:social-welfare}. An aspect of prosociality \revisedb{as we apply it here} is that \revisedb{the agents perform actions that benefit others even if suboptimal to themselves.}\revisedbst{because the agents look out for each other's benefit,} \revisedb{Therefore, in our approach the agents who are worse off are likelier to benefit from prosociality of others than the opposite, and hence} fewer agents are poorly off, and the disparity in outcomes across the agents reduces.

\item [Social welfare] captures the idea of how much a society (as a whole) gains. Here, we specifically identify how a particular approach to coordination resource usage may lead to aggregate outcomes. This point can be framed as a question about the feasibility of a decentralized approach in producing social welfare relative to an approach that is fully centralized or incorporates elements of centralization through a distinguished entity to help coordinate.

\end{description}


\subsection{Significance and Novelty}
This paper synthesizes two important perspectives on prosociality. First, the individual decision-making perspective incorporates a notion of guilt based on inequity aversion \cite{Fehr1999}, which posits that people may be self-interested, but their decisions are affected by how relatively poorly others fare. In a multiagent system, a prosocial agent in \framework would maintain some awareness of the outcomes received by other agents. When those outcomes are especially low for another agent, the first agent would notice the inequity and may feel guilt or exhibit aversion to that inequity. The agent would take decisions that lead to a reduction in that observed inequity.

Second, the societal perspective on prosociality is based on Rawls' \shortcite{Rawls-99:justice} landmark theory of justice that focuses on designing a just society. This perspective is broadly supported by more recent works as well, for example, by \citet{Adler-19:social-welfare}.
Specifically, \framework supports Rawls' doctrine of improving the outcome for whoever is the worst off. In other words, this doctrine advocates that we do not maximize throughput if doing so leads to some agents starving. We adopt the Maximin formulation of Rawls' doctrine  as a basis to measure fairness. By bringing in guilt, we can support norms that emerge in a bottom-up and adaptive manner without needing a central society enforcer.

\subsection{Organization}
Section~\ref{sec:Framework} details the \framework framework.
Section~\ref{sec:prosociality} describes how we model prosociality.
Section~\ref{sec:casestudy} describes a simulated traffic
intersection for evaluation. Section~\ref{sec:result} discusses our
results. Section~\ref{sec:concl} provides a summary of our
contributions and an outlook for future work.

\section{The \framework Normative Framework}
\label{sec:Framework}

\begin{figure}[htb]
  \begin{centering}
  \includegraphics[width=0.5\columnwidth]{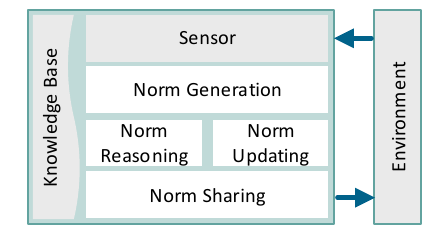}
  \caption{Components of a \framework agent. Each \framework agent has sensors via which it perceives the environment, a knowledge base consisting of a set of norms (and norm structures), and components for generating, reasoning, updating, and sharing norms.}
  \label{fig:cha-model}
  \end{centering}
\end{figure}

Figure~\ref{fig:cha-model} depicts a schematic representation of an agent in \framework 
and different phases of its normative life cycle. 
Algorithm~\ref{alg:the_alg} outlines the decision loop for a \framework agent in four phases enacted via the agent's corresponding components: \emph{Norm Generation}, \emph{Norm Reasoning}, \emph{Norm Updating}, and \emph{Norm Sharing} (shown with labels in Algorithm~\ref{alg:the_alg}). 
In this section, we describe norm representation in \framework and explain each phase of its normative life cycle along with the associated pseudocode that demonstrates the dynamics of \framework. 

\subsection{Norm Representation}
As explained above, a norm in \framework is \emph{regulative}: it characterizes how the agents ought to interact in specific situations. A ``norm structure'' is either a norm or a precursor to a norm that is based on a deontic operator \revised{proposed by} \citet{Garcia2009}. We adopt a continuous notion of deontics \cite{Frantz2013}, ranging from \emph{prh} to \emph{obl} with \emph{may} in the middle. Initially, a norm structure is neutral. We express its neutrality using the operator \emph{may}. The idea is that, as interactions take place, a norm structure may strengthen in one or the other direction or weaken back toward neutral. In cases of interest, a norm structure would strengthen enough in one direction or the other. Specifically, we say a norm emerges when the operator strengthens to \emph{obl} or \emph{prh}. Table~\ref{tab:syntax} gives \framework's syntax.

\begin{table}[!htb]
\centering
\caption{Syntax of a norm structure. Antecedent is a condition on the system state; Consequent is a deontic operator applied on an action.}
\label{tab:syntax}
\begin{tabular}{@{ }l@{ }l@{ }l}\toprule
Norm & ::= & $\langle$Antecedent, Consequent$\rangle$ \\
Antecedent & ::= & Condition \\
Consequent & ::= & Operator(Action) \\
Operator & ::= & \emph{may} $\mid$ \emph{obl} $\mid$ \emph{prh}\\
\bottomrule
\end{tabular}
\end{table}

\revised{In \framework,} given a state as it perceives, an agent applies any norm structure whose antecedent is true in that state \revised{(\emph{Norm Generation})}. 
It performs the action in the norm if its deontic is \emph{obl}, does not perform it if the deontic is \emph{prh}, and chooses either for \emph{may} \revised{(\emph{Norm Reasoning})}. 
As the agent gains experience, \emph{obl} or \emph{prh} may begin to dominate, indicating a norm being learned \revised(\emph{Norm Updating}). 
\revised{Sections~\ref{sec:norm-generation}--\ref{sec:norm-sharing} describe these phases of normative lifecycle in detail.}

\subsection{Norm Generation Phase}
\label{sec:norm-generation}
An agent \emph{perceives} its environment through its sensors and receives a local \emph{view} (Line~2). A local \emph{view} is a snapshot of the system from the agent's perspective. An agent's \emph{view} is in general partial. The sensors are potentially specialized to each domain: a sensor is a domain-dependent function to map the state of the environment to a description of the state in the agent's representation. We express an agent's \emph{views} using predicates over the state of the system. For example, in a traffic scenario, a possible \emph{view} is the traffic in one part of the road that identifies positions and directions of movement of the other vehicles in reference to the agent's vehicle. 

For each \emph{view}, the agent checks whether a conflict would happen using the function $f_{\text{conflict}}$ based on the possible next states, relying on domain information to determine which outcomes are undesirable. The agent gets the \emph{view} $v_{t}$---the agent's view at time $t$, that would lead to conflict, and \emph{conflictAgent} (Line~3). 
\emph{conflictAgent} is the other agent involved in the conflicting \emph{view} $v_{t}$.

Here \emph{normSet} is the set of norms represented by the agent
(Line~4). A norm applies to an agent if its antecedent matches the agent's
current view. If no norm in normSet is applicable (Line~5), the agent
generates a norm structure based on the current view ($v_{t}$) as its
antecedent, initially with a \emph{may} operator applied to the
\emph{action} that would lead to the conflict (Line~6). The generated
norm structure is added to normSet (Line~7).

\begin{algorithm}
  \caption{Decision loop for a \framework agent}
  \label{alg:the_alg}
  \begin{algorithmic}[1]
  \While {True}{

    \textbf{\# Norm Generation:}
    \State \text{\textit{view}} $ \gets \text{perceiveEnvironment}(\textit{Sensors})$;
     \State $v_{t},\text{conflictAgent}\gets \text{conflictDetect}(\textit{view}, f_{\text{conflict}})$;
     \State $\textit{normSet}$ $ \gets \text{current norms}$;

        \If {$\forall \langle \text{ant}, \text{con}\rangle \in \textit{normSet}\colon v_t\not\vdash \text{ant}$}
           \State $\text{\textit{n}}\gets \langle v_t, \text{may}(\textit{action})\rangle$;
                \State $\textit{normSet}$ $\gets \text{\textit{normSet}}  \cup \{\textit{n}\}$;
        \EndIf

    \textbf{\# Reasoning:}
          \State $\text{\textit{n}}\gets \text{getApplicableNorm}(\textit{normSet})$;
     \State $\text{\textit{a}}\gets \text{actionSelection}(\textit{n, $\epsilon$-greedy})$;
     \State $\textit{execute}\left(\textit{a}\right)$;
    \State $a_{c}$ $ \gets \text{perceiveAction}(\text{conflictAgent}, \textit{Sensors})$;

    \textbf{\# Updating:}
     \State $\text{\textit{r}}\gets \text{jointActionReward}(\textit{Payoffs(a,$a_c$)})$;

     \State $\text{\textit{U(n,t,a)}}\gets
     \textit{updateUtility}\left(\textit{n, t, a}\right)$;

    \textbf{\# Sharing:}
       \State \textit{shareExperience()};
  \EndWhile}
  \end{algorithmic}
\end{algorithm}

\subsection{Reasoning Phase}
\label{sec:norm-reasoning}
The agent retrieves an applicable norm; selects an action according to
its \revised{learning algorithm} and an applicable norm, and performs the action
(Lines~8--10). Next, the agent senses its conflicting agent's action,
$a_c$ (Line~11).

\framework agents apply reinforcement learning with the $\epsilon$-greedy strategy for exploration and exploitation of the environment \revised{ \cite{Sutton+Barto-18:RL}}. The
$\epsilon$-greedy approach offers two choices for each agent---select
a random action (Exploration, with probability $\epsilon$), or follow
the applicable norm (Exploitation, with probability $1-\epsilon$). An agent estimates $\epsilon$ via an exponential function
($e^{-Em}$), where $E$ is a constant, and $m$ is the number of times
that the same situation has arisen before. Consequently, $\epsilon$ is
high early (more exploration) and low later (more exploitation). \revised{Decaying exploration has been considered widely in the Reinforcement Learning literature \cite{Singh2004ConvergenceRF}. The decay method is traditionally a linear or exponential decay. Exponential decay has been shown to lead to faster convergence \cite{Avraham2008}.}

\subsection{Updating Phase}
\label{sec:norm-updating}
The agent updates its utility for the applied norm. After reasoning and selecting an action, based on the joint action of its conflicting agent, it receives a reward according to a payoff matrix (Line~12) and updates its utility via Equation~\ref{utility} (Line~13).
\begin{equation} \label{utility}
U(n,t,a)= (1-\alpha)\times U(n,t-1,a)+\alpha\times r(n,t,a).
\end{equation}
Here, $U(n,t,a)$ and $U(n,t-1,a)$ stand for the utility of following
norm structure $n$ by performing action $a$ at times $t$ and $t-1$,
respectively; $0\leq \alpha\leq 1$ is a parameter to trade off
exploitation with exploration, and $r(n,t,a)$ indicates the reward
based on a payoff matrix (examples in Section~\ref{sec:casestudy}).

Each agent assigns two utility values to each norm structure for when the norm structure is, respectively, fulfilled (followed) and violated.
Note that following or violating the norm structure eventually would lead to one of the operators with normative force, \emph{obl} or \emph{prh}, respectively, depending on whichever has the higher utility.

\subsection{Sharing Phase}
\label{sec:norm-sharing}
In \framework, each agent passes on its \emph{experience}  to
incoming members of the same \emph{type} (Line~14). Here, the experience is given by the
utilities associated with different states and action outcomes and agents of the same type are those with the same
goals. Thus, an incoming member obtains experience from
members of the same type who have some experience to share. This approach
fits in with technologies such as Vehicle-to-vehicle (V2V)
communication. We assume that agents pass on their experience
truthfully; false information \cite{Staab2008} is out of our scope.

\section{Acting Prosocially in \framework}
\label{sec:prosociality}

The above approach may lead to unfair outcomes. For example, imagine that in a road traffic scenario, agent $i$ has the right of way, and agent $j$ (conflicting with agent $i$) must stop. In other words, the norm for agent $i$ is initially \emph{obl}(\emph{Go}) and the norm for agent $j$ is \emph{prh}(\emph{Go}). If there is heavy traffic in agent $i$'s direction, agent $j$ may have to wait for an arbitrarily long time. Long delays for some agents but not for others indicates unfairness.

In \framework, the agents act prosocially through inequity aversion \cite{Fehr1999}. An agent incorporates another's cumulative costs in its utility to help the latter. That is, instead of following an applicable norm that would benefit it, 
the agent performs an action that benefits the other party. In the traffic domain, such an action might be to wait and let another vehicle cross when you truly have the right of way.
A benefit of using the cumulative cost from when an agent enters the system until it gets the opportunity to benefit from the resource is that it can help prevent starvation. We assume that agents know about each other's costs to be able to accommodate them in their decision making. 
\revised{This assumption is consistent with real-world scenarios, in which people need to be aware of others' condition first, to act prosocial. In human societies, people can figure out each other's valuations enough to develop a sense of prosociality \cite{SHAW201440}. For artificial agents, this information must be either hardwired or explicitly communicated by the agents \cite{Hao2016}.}

We incorporate \emph{guilt} \cite{Lorini2015} as a \emph{guilt disutility} to realize concessions. 
\revised{Please note that ``guilt'' here is not meant to capture psychosocial realism but is a convenient metaphor for a feeling that leads to prosocial behavior.}
\revised{Here, $\delta_{i, j}$, as computed in Equation~\ref{Guilt_disuti}, expresses the guilt perceived by agent $i$ with respect to $j$ in state $s$.}

\begin{equation}\label{Guilt_disuti}
\delta_{i, j}(s)=
-\beta_{i} \left(\max \left(f_{j}(s)-f_{i}(s)-\emph{c}, 0\right)\right)\!.
\end{equation}
Here, $f_x(s)$ computes the total cost paid by agent $x$ until the present. An example of the calculation of $f_x(s)$ is discussed in Section~\ref{sec:casestudy} (Equation~\ref{payoff}). 
Agent $i$'s propensity toward guilt is captured in $\beta_{i}$: $\beta_i=0$ means no guilt, and $\beta_i=1$ means maximal guilt. Here, $c$ is the threshold of inequity at which guilt kicks in.

Algorithm~\ref{alg:the_fair__alg} adds prosociality to Algorithm~\ref{alg:the_alg} on Line~9 (actionSelection). For simplicity, we enable prosocial learning once the system has converged---otherwise, the system may never stabilize, especially with high values of $\beta_{i}$. Note that convergence is defined in Section~\ref{sec:Simulation_setup}.
$U_i^P$, the utility incorporating prosociality is initialized to the converged utility (Lines~2--4). Below, $\neg n$ is the complement of norm $n$: it has the same antecedent but \emph{obl} instead of \emph{prh} or vice versa. That is, $\neg\langle p, obl(q)\rangle = \langle p, prh(q)\rangle$. Here, $n$ is preferred by agent $i$ and $\neg n$ by its conflicting agent $j$.

Agent $i$ receives agent $j$'s cost ($f_j(s)$, \revisedbst{Line~6} \revisedb{Line~5}). If the difference in costs is below constant $c$, $i$ follows norm $n$ (\revisedbst{Line~14} \revisedb{Line~17}). Otherwise, if agent $i$ has not learned to concede ($U_i^P(n,t,a) > U_i^P(\neg n,t,a)$) (\revisedbst{Line~7}\revisedb{Line~6}), $i$ follows norm $n$ (\revisedbst{Line~8}\revisedb{Line~11}), adds its guilt disutility (a negative value) to its prosocial utility (\revisedbst{Lines~9--10}\revisedb{Lines~8--9}). Eventually, $U_i^P(n,t,a)$ would fall and agent $i$ concedes to agent $j$ (\revisedbst{Line~12}\revisedb{Line~14}).

\begin{algorithm}
  \caption{Prosocial learning strategy for agent $i$}
  \label{alg:the_fair__alg}
  \begin{algorithmic}[1]
  \If {{Converged}}
  \If {$U_i^P$ \emph{Not Initialized}}
   \State  \ \ \ \ \  $U_i^P(n,t,a) \gets U_{i}(n,t,a)$
      \State  \ \ \ \ \ $U_i^P(\neg n, t, a) \gets U_{i}(\neg n, t, a)$
    \EndIf
     \State \emph{Receive} agent $j$'s total cost, $f_{j}(s)$ via Sensors
     \If {$f_{j}(s)-f_{i}(s) > c$ \fbf{and} 
      $U_i^P(n,t,a) > U_i^P(\neg n,t,a)$}
      \State \revisedbst{\emph{Follow} norm $n$}
     \State  $\delta_{i, j}=-\beta_{i} \left(\max \left(f_{j}(s)-f_{i}(s)-c,0 \right)\right)$
      \State $U_i^P(n,t,a) = U_i^P(n, t-1, a)+\delta_{i, j}$
    \State \# \emph{Follow} norm $n$
    \State \revisedb{ $a \gets \text{actionSelection}(n, \epsilon=0)$}
     \ElsIf {$f_{j}(s)-f_{i}(s) > c$ \fbf{and} 
     $U_i^P(n,t,a) < U_i^P(\neg n,t,a) $}
     \State \# \emph{Concede} to agent $j$ and \emph{follow} norm $ \neg n$
     \State \revisedb{ $a \gets \text{actionSelection}(\neg n, \epsilon=0)$}
      \Else {}
      \State \# \emph{Follow} norm $n$
      \State \revisedb{ $a \gets \text{actionSelection}(n, \epsilon=0)$}
      \EndIf

  \Else
    \State $a \gets \text{actionSelection}(n, \epsilon\text{-greedy})$

  \EndIf
  \end{algorithmic}
\end{algorithm}


\section{Traffic Setting as Illustration}
\label{sec:casestudy}

We evaluate \framework in a simulated intersection understood as a multiagent system, where vehicle agents continually arrive and depart. We select this setting because it is powerful enough to illustrate \framework but not so powerful that details of the setting overwhelm our main points. Related works, for example, \citet{IJCAI-16:Silk} and \citet{Morales:2015}, adopt similar settings as in this paper. 
Please note that our focus is not to model the complexities of real-world traffic but to demonstrate the working of \framework. \framework could be potentially applied in more complex settings, for example, with a network of intersections.

\begin{figure}[htb]
  \begin{centering}
  \includegraphics[width=\columnwidth]{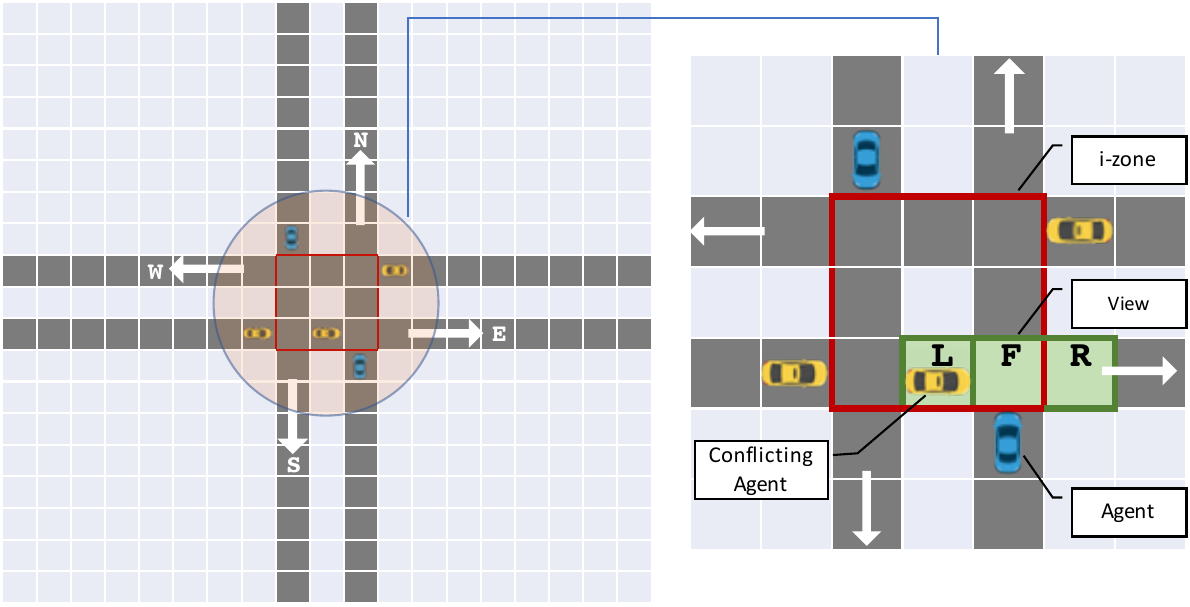}
  \caption{The modeled intersection mapped to a grid of cells. Traffic flows in four directions (northbound, southbound, westbound, and eastbound). We consider the possibility of conflict within the i-zone. The i-zone is composed of eight cells in the middle (excluding the middle grey cell). From the perspective of the car specified as the Agent (shown in the bottom right of the inset picture), its view is composed of three cells that it must observe to detect and avoid conflicts. L refers to the cell to its left, F to the cell directly to its front, and R to the cell to its right. The car specified as Conflicting Agent is in conflict with the Agent.}
  \label{fig:intersection-center}
  \end{centering}
\end{figure}

As Figure~\ref{fig:intersection-center} illustrates, we map an intersection and its vicinity to a grid. Traffic flows in four directions (north, south, west, and east). The \emph{intersection zone (i-zone)} in the middle is composed of eight cells, highlighted in Figure~\ref{fig:intersection-center}. The light grey cells, including the center, are to be ignored. Cars travel along the grid at the speed of one cell per time tick. A vehicle may continue \revisedb{moving} on a straight path or may randomly turn left or right in the i-zone. Agents of the same \emph{type} are those traveling in the same direction, for example, all vehicles that are traveling east.

A conflict arises when two \emph{conflicting agents} are about to occupy the same i-zone cell together. Conflicting agents observe each other's actions without access to each other's internal policies (Algorithm~\ref{alg:the_alg} Line~11). If there is no norm in a conflicting situation, each agent creates a norm structure using the neutral operator, \emph{may}. As in Section~\ref{sec:Framework}, a norm emerges as the agents gain experience and the deontic operator becomes stronger, namely, \emph{obl} or \emph{prh}.

The antecedent of a norm refers to the values of three cells (Left $L$, Front $F$, Right $R$) identified in Figure~\ref{fig:intersection-center} with respect to the vehicle (agent) entering from the bottom of the intersection. $L$, $F$, and $R$ constitute the view of that vehicle. Our grammar can specify cells with one of these six values: $\goR$ (vehicle heading east), $\goL$ (vehicle heading west), $\goD$ (vehicle heading south), $\goU$ (vehicle heading north), $\gono$ (empty), and $\go$ (any of four directions). An example norm, $L(\goR)$ = \emph{prh}(\emph{Go}), means that an agent is prohibited from proceeding if it observes a vehicle in cell $L$ heading east ($\goR$). Figure~\ref{fig:intersection-center} shows a similar case.

Table~\ref{tab:payoff-matrix} shows our payoff matrix, which
represents a \emph{social dilemma} game, with the added twist of dynamism.
Its constants are typical \cite{Airiau2014,sugawara-AAMAS14-emergence}; the unselfishness term (that is, $u$) is novel to \framework. 
\revised{In a single round,} the best positive payoff refers to the situation where one agent chooses a selfish action (\emph{Go}) and the conflicting agent chooses the unselfish action (\emph{Stop}). 
\revised{However, in a multi-round setting where the game is repeated, if the value of $u$ is low, it is more rewarding to be cooperative than to be selfish, and if the value of $u$ is high, it is more rewarding to be selfish than to be cooperative. 
Our flexible payoff matrix allows for the representation of various societies of agents.}

\begin{table}[!htb]
\centering
\caption{Payoff matrix ($u$ is the unselfishnessCost).}
\label{tab:payoff-matrix}
\setlength{\extrarowheight}{2pt}
\begin{tabular}{cc|c|c|}
  & \multicolumn{1}{c}{} & \multicolumn{2}{c}{Agent $j$}\\
  & \multicolumn{1}{c}{} & \multicolumn{1}{c}{$\emph{Go}$}  & \multicolumn{1}{c}{$\emph{Stop}$} \\\cline{3-4}
  \multirow{2}*{Agent $i$}  & \emph{Go} & $-6.0$, $-6.0$ & $5.0$, $u$ \\\cline{3-4}
  & \emph{Stop} & $u$, $5.0$ & $u$, $u$ \\\cline{3-4}
\end{tabular}

\end{table}

The worst negative payoff relates to the situation when both agents are selfish, meaning that both decide to \emph{Go}, thereby causing a collision. Equation~\ref{payoff} defines the payoff of an unselfish action for agent $x$ in state $s$, that is, $f_x(s)$ in Section~\ref{sec:prosociality}. The $\max$ ensures that the payoff of stopping is never worse than of a collision, \revised{and as a result the value of unselfish action ranges from $0$ to $-6$.}

\begin{equation}\label{payoff}
\textit{unselfishnessCost}_x(s) = \max(-d_x(s)^p,-6)
\end{equation}
\revised{We use a power function to calculate the delay.} Here, $d_x$, the delay experienced by agent $x$, equals the number of
ticks that $x$ must stop before entering the i-zone. The cost
increases exponentially with delay. Specifically, we require the exponent of the delay term to be greater than 1, that is, $p>1$. \revised{Intuitively, when the delay (for a specific agent) is high, the penalty that the agent assigns to stopping should be higher. Equation~\ref{payoff} has this property.}

\section{Results: Testing the Hypotheses}
\label{sec:result}




\revised{To evaluate \framework in the traffic setting detailed in Section~\ref{sec:casestudy}, we propose the following hypotheses based on the four themes --- \fsl{efficient resolution of conflicts}, \fsl{dynamic adaptation}, \fsl{fairness}, and \fsl{social welfare} --- that we motivate in Section~\ref{sec:introduction}.
For brevity, we omit the corresponding null hypotheses indicating there are no gains under \framework.
}


\begin{itemize}
\item \revised{Hypothesis H\fsub{efficient}: \framework leads to the emergence of norms that resolve conflicts while improving system-level performance outcomes.}

\item \revised{Hypothesis H\fsub{dynamic}: \framework leads to the emergence of norms that can adapt based on changes to the environment. That is, an emerged norm can fade out once the environment changes and be replaced by a new norm.}

\item \revised{Hypothesis H\fsub{fairness}: \framework leads to the emergence of norms under which fairness is achieved. Specifically\revisedbst{, a variant} 
\revisedb{in \framework,} the agents observe the outcomes of others and act toward the benefit of those poorly off, leads to improved fairness.}

\item \revised{Hypothesis H\fsub{social}: \framework leads to the emergence of norms that yield higher societal gains than both a representative central and a representative hybrid approach to sharing resources.}

\end{itemize}

\revised{
We now describe our simulation setup to evaluate \framework and discuss each of the hypotheses, that is, H\fsub{conflict}, H\fsub{efficient}, H\fsub{dynamic}, H\fsub{fairness}, and H\fsub{social}, in greater detail.
}

\subsection{Simulation Setup}
\label{sec:Simulation_setup}

We model a traffic intersection (19 cells per lane: 72 cells in all with eight cells in the i-zone) as an environment in Repast \cite{North2013}.

Our results are averaged over 1,000 trials. Convergence is considered to happen if the utilities associated with the norms converge to within $10^{-3}$ (our \emph{convergence parameter}). 

We set the initial utility values to 0 at $t=0$, $\alpha=0.2$ in Equation~\ref{utility} \cite{Morales:2015}. We set $E=0.05$ in the exponential function ($e^{-Em}$) used to set the exploration probability in the $\epsilon$-greedy exploration approach in the reasoning phase \cite{Tantawy}. 

\revisedb{The impact of delay, that is, $p$, the exponent of the delay term in Equation~\ref{payoff} is set to 1.1 for all agents.}
We choose \revisedbst{these values} \revisedb{this value} somewhat arbitrarily because adjusting \revisedbst{them} \revisedb{it} would affect only the rate of convergence in our experiments, not whether convergence takes place.

For the fairness experiment to test Hypothesis H\fsub{fairness}, we need to set the following parameters. 
\revisedst{We choose these values somewhat arbitrarily and adjusting them would affect only the rate of convergence in our experiments, not whether convergence takes place.}

\begin{itemize}
\item The propensity or weight of guilt, that is, $\beta$ in Equation~\ref{Guilt_disuti}. Note that $\beta=0$ means no guilt \revised{and leads to no fairness} and $\beta=1$ means maximal guilt\revised{, which leads to repeated yielding by agents in favor of others}. For simplicity of evaluation, we set the same value of $\beta$, specifically, $\beta = 0.5$, for all agents.

\item The threshold of inequity tolerated by the agents, that is, $c$ in Equation~\ref{Guilt_disuti}. For simplicity, we set the same value of $c$, specifically, $c = 4$, for all agents.

Note that the fairness parameters' values \revisedb{(i.e., $c$ and $\beta$)} do not affect the convergence, since as mentioned in Section~\ref{sec:prosociality}, we enable prosocial learning once the system has converged---otherwise, the system may never stabilize, especially with high values of beta. Fairness parameters' values affect the rate at which the agents become prosocial.

\end{itemize}

\subsection{Testing the Efficient Resolution Hypothesis}
\label{sec:efficient}
Hypothesis H\fsub{efficient} states that emergent norms improve system-level goals---yield lower total average delays across an intersection.
\revised{The corresponding null hypotheses indicate that there are no improvements in the system-level goals under \framework.}

We consider a static setting in which we fix the traffic flows for the north-south and east-west directions. We first set the traffic flow distributions as the same for north-south and east-west directions. We observe that approximately half of the time (508 out of 1,000 simulation runs), north-south vehicles learn to Go, and east-west vehicles learn to Stop in conflict situations. In the remaining 492 times, the reverse norm arises. The population converges to one or the other norm depending on whether Go or Stop is more common for east-west than for north-south vehicles and thus minimizes collisions.

\begin{figure}[!htb]
    \centering
    \begin{tikzpicture}
    \begin{axis}[
    title={},
    height=6cm, 
    width=8cm,
    xlabel={Time in 1,000 ticks},
    ylabel={Collisions},
    xmin=0,xmax=25,
    ymin=-25,ymax=270,
    ]
     \addplot +[mark=none,] table [x=tick, y=collisions, col sep=comma]
     {data/collision_counts_window-1000.csv};
     \addplot +[mark=none,black, dashed] coordinates {(0,0) (25, 0)};
    \end{axis}
    \end{tikzpicture}
    \caption{Total number of collisions per 1,000 ticks.
    }
    \label{fig:collisions}
\end{figure}
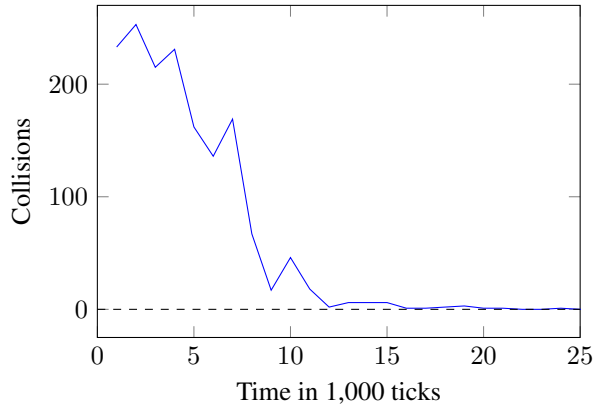

Our next setting is also static but with unequal traffic: the north-south orientation has a (30\%) higher traffic volume than the east-west orientation. 
Figure~\ref{fig:collisions} shows the total number of collisions per 1,000 ticks. Since there are four cells in the simulated intersection that have the potential of conflict, the maximum number of collisions is four per tick.
After about 20,000 ticks, the number of collisions decreases dramatically. 
After 25,000 ticks (not shown here for brevity), the changes in the average utility converge to within $10^{-3}$, our convergence parameter. 
\revisedst{This trend in convergence is similar to that observed in Figures 4 and 5}

Based on the asymptotic reduction in collisions and the convergence in utilities,
we conclude that vehicle agents have learned new norms to
avoid collision.
Table~\ref{tab:emerged-norms-static} shows the emergent norms: east-west vehicles learn to Stop in conflicting
situations with north-south vehicles. Since the north-south orientation has higher traffic volume, the converged norms are efficient---average delay is lower when vehicles in the direction with the lower volume
Stop and in the direction with the higher volume Go, than the other way around. Norms emerged in this experiment provide evidence to support Hypothesis H\fsub{efficient}.

\newcommand{\tand}{\!\land\!}
\begin{table}[!htbp]
\caption{Emerged norms for static setting with equal and unequal traffic for the first half and all the time, respectively.}
\label{tab:emerged-norms-static}
\centering

\begin{tabular}{p{3cm} c c}
\toprule
\textbf{} & \textbf{Precondition} & \textbf{Modality} \\ \midrule

\multirow {2}{3cm}{\emph{Eastbound and }\emph{Westbound}} & $L(\go)\tand F(\go)\tand R(\goL)$& $\mathit{prh}(\text{Go})$\\ \cmidrule{2-3}
& $L(\goR)\tand F(\go)\tand R(\go)$& $\mathit{prh}(\text{Go})$\\ \midrule
\multirow {2}{3cm}{\emph{Southbound and }\emph{Northbound}} & $L(\go)\tand F(\go)\tand R(\goL)$& $\mathit{obl}(\text{Go})$\\ \cmidrule{2-3}
& $L(\goR)\tand F(\go)\tand R(\go)$& $\mathit{obl}(\text{Go})$\\ \bottomrule
\end{tabular}
\end{table}

\subsection{Testing Dynamic Adaptation Hypothesis}
\label{sec:Dynamic}
%
The Hypothesis H\fsub{dynamic} states that \framework adapts to environmental changes---changes in traffic flow through an intersection.
\revised{The corresponding null hypothesis indicates that \framework does not adapt to environmental changes.}
In this experiment, we start with a fixed traffic flow distribution where the north-south orientation has 30\% higher traffic than the east-west orientation, let it converge, and then reverse the pattern (that is, the east-west orientation ends up with 30\% more traffic). Doing so helps us determine whether norms learned in one setting persist when the traffic changes.

To test Hypothesis H\fsub{dynamic}, we (1) measure root mean square deviation (RMSD) of average utility in a sliding window of 1,000 ticks, computed as
RMSD\fsub{t} = $\sqrt{\dfrac{\Sigma_{t=1}^{n} (x_i - \bar{x})^2}{n-1}}$, where $t$ is the current tick; $x_t$ is the utility in the current tick; $\bar{x}$ is the average utility in the current sliding window; and $n$ is the window size, and (2) perform a two-sample Kolmogorov-Smirnov (KS) test on successive sliding windows. 

\begin{figure}[!htb]
    \centering
    \begin{tikzpicture}
    \begin{axis}[
    title={},
    height=6cm, 
    width=8cm,
    xlabel={Time in 1,000 ticks},
    ylabel={Utility},
    xmin=0,xmax=65,
    ymin=-8,ymax=8,
    ]
     \addplot +[mark=none,dashed] table [x=tick, y=goutility, col sep=comma]
     {data/west_utility_window-1000.csv} node[above,pos=0.84]{Go};
     \addplot +[mark=none,] table [x=tick, y=stoputility, col sep=comma]
     {data/west_utility_window-1000.csv} node[above,pos=0.82]{Stop};
     \node[fit={(25,-2.5) (43,0.5)}, inner sep=0pt, draw=red!50!black, dashed, fill=red, opacity=0.1] (rect) {};
     \node[fit={(50,3.5) (64.5,6.5)}, inner sep=0pt, draw=blue!50!black, dashed, fill=blue, opacity=0.1] (rect) {};
    \end{axis}
    \end{tikzpicture}
    \caption{Utilities of westbound vehicles for the dynamic setting. The reported figures are averaged over a window size of 1,000 ticks.}
    \label{fig:west-utility}
\end{figure}
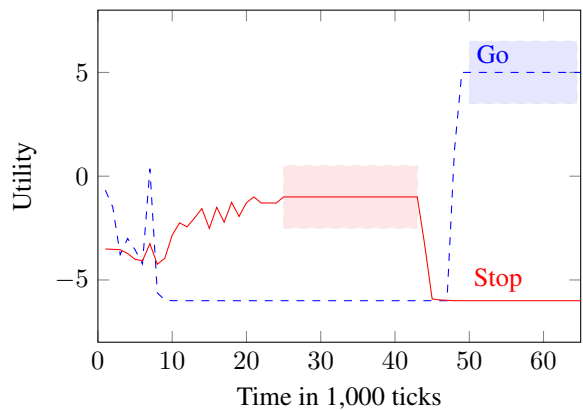

Figure~\ref{fig:west-utility} shows the change in average utility (for a sliding window of 1,000 ticks) for westbound vehicles. By $t\approx \text{25,000}$ (RMSD\fsub{25,000} = 0 for east-west vehicles Stop; $p<0.01$), the norm learned by east-west vehicles is to Stop in case of conflict, just as in Table~\ref{tab:emerged-norms-static} and Figure~\ref{fig:collisions}. We highlight this norm in Figure~\ref{fig:west-utility} with the shaded box in the middle.

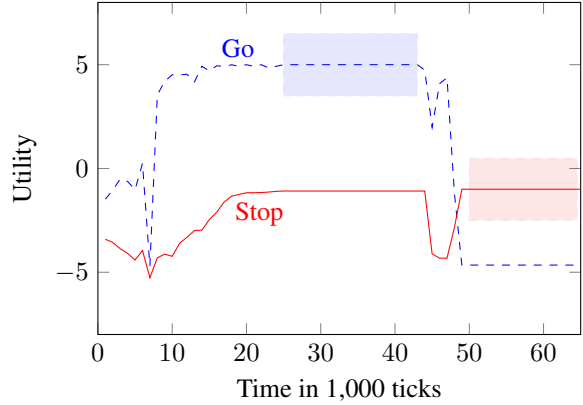
\begin{figure}[!htb]
    \centering
    \begin{tikzpicture}
    \begin{axis}[
    title={},
    height=6cm, 
    width=8cm,
    xlabel={Time in 1,000 ticks},
    ylabel={Utility},
    xmin=0,xmax=65,
    ymin=-8,ymax=8,
    ]
     \addplot +[mark=none,dashed] table [x=tick, y=goutility, col sep=comma]
     {data/south_utility_window-1000.csv} node[above,pos=0.34]{Go};
     \addplot +[mark=none,] table [x=tick, y=stoputility, col sep=comma]
     {data/south_utility_window-1000.csv} node[below,pos=0.31]{Stop};
     \node[fit={(25,3.5) (43,6.5)}, inner sep=0pt, draw=blue!50!black, dashed, fill=blue, opacity=0.1] (rect) {};
     \node[fit={(50,-2.5) (64.5,0.5)}, inner sep=0pt, draw=red!50!black, dashed, fill=red, opacity=0.1] (rect) {};
    \end{axis}
    \end{tikzpicture}
    \caption{Utilities of southbound vehicles for the dynamic setting. The reported figures are averaged over a window size of 1,000 ticks.}
    \label{fig:south-utility}
\end{figure}

After $t= \text{44,000}$, the traffic pattern is reversed, and by $t\approx \text{50,000}$ (RMSD\fsub{50,000} = 0 for east-west vehicles Go; $p<0.01$), the new norm is for east-west vehicles to Go in case of conflict. We highlight this norm in Figures~\ref{fig:west-utility} and~\ref{fig:south-utility} with boxes in the right parts. Eastbound and northbound vehicles have the same outcomes as westbound and southbound vehicles, respectively.

Table~\ref{tab:emerged-norms-dynamic} shows how norms change when we reverse the traffic pattern, thus supporting Hypothesis H\fsub{dynamic}.

\begin{table}[!htbp]
\caption{The norms as emerged in the dynamic setting when the traffic flow has been reversed.}
\label{tab:emerged-norms-dynamic}
\centering

\begin{tabular}{p{3cm} c c}\toprule
\textbf{} & \textbf{Precondition} & \textbf{Modality} \\ \midrule

\multirow {2}{3cm}{\emph{Eastbound and }\emph{Westbound}} & $L(\go)\tand F(\go)\tand R(\goL)$& $\mathit{obl}(\text{Go})$\\ \cmidrule{2-3}
& $L(\goR)\tand F(\go)\tand R(\go)$& $\mathit{obl}(\text{Go})$\\ \midrule
\multirow {2}{3cm}{\emph{Southbound and }\emph{Northbound}} & $L(\go)\tand F(\go)\tand R(\goL)$& $\mathit{prh}(\text{Go})$\\ \cmidrule{2-3}
& $L(\goR)\tand F(\go)\tand R(\go)$& $\mathit{prh}(\text{Go})$\\ \bottomrule
\end{tabular}
\end{table}


\revisedb{Note that, given that the agents have developed a notion of norms already, they tend to adapt faster to environmental changes, that is it takes $t\approx \text{25,000}$ ticks to initially learn the norms, and $t\approx \text{6,000}$ ticks to adapt to new environmental changes and develop new norms. If the flows of cars invert intermittently (e.g., every x ticks), as long as the final traffic difference is significant (e.g., 30\%), we still would expect \framework to be adaptive, however, it could take longer to converge.}

\subsection{Testing the Fairness Hypothesis}

\label{sec:Fairness_Hypothesis}
%
Hypothesis H\fsub{fairness} concerns disparities in resource allocation---excessive delays for some vehicles to enter the intersection while others proceed quickly.
We define a fair society as one that supports the Maximin criterion \cite[p.~153]{Rawls-99:justice}. In the present setting, a fair society means one in which no agent is deprived of resources for long periods of time. We understand fairness (or the absence of fairness) as an outcome of different norms.

Consider the prosocial learning strategy given in Algorithm~\ref{alg:the_fair__alg}. As in Section~\ref{sec:efficient} (unequal traffic setting), we set north-south flows to have 30\% more traffic than east-west flows. We saw that east-west vehicles learned to Stop in conflicting situations with north-south vehicles. Now, we verify whether a north-south vehicle can act prosocially in a conflicting situation by yielding to an east-west vehicle if it experiences a delay above a certain threshold.

\begin{figure}[!htb]
    \centering
    \begin{tikzpicture}
    \begin{axis}[
    scaled ticks=false,
    tick label style={/pgf/number format/fixed},
    title={},
    height=6cm, 
    width=8cm,
    xlabel={Time in 1,000 ticks},
    ylabel={Utility},
    xtick={35000,40000,45000,50000,55000,60000},
    xticklabels={35,40,45,50,55,60},
    xmin=35000,xmax=60000,
    ymin=-8,ymax=8,
    ]
     \addplot +[mark=none,dashed] table [x=tick, y=go, col sep=comma]
     {data/southbound_fair.csv} node[above,pos=0.15]{Go};
     \addplot +[mark=none,] table [x=tick, y=stop, col sep=comma]
     {data/southbound_fair.csv} node[above,pos=0.15]{Stop};
    \end{axis}
    \end{tikzpicture}
    \caption{Prosocial utilities of southbound vehicles.
    }
    \label{fig:southbound}
\end{figure}
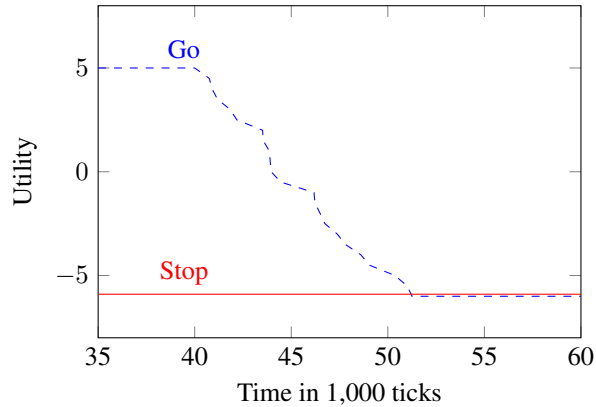

Figure~\ref{fig:southbound} shows the change in the average prosocial utility ($U_i^P$) of southbound vehicles. Prosocial learning can be activated after convergence; we activated it at $t= \text{40,000}$. By $t\approx \text{51,000}$, the agents have learned to be prosocial. Northbound vehicles (figure omitted for space) show the same trend as southbound vehicles.

Below, \revisedbst{\framework} \revisedb{\frameworkNeu} refers to the variant form \revisedb{without prosociality, i.e., without incorporating the guilt disutility}. 
We evaluate \frameworkNeu and \framework's performance in terms of delays. Table~\ref{tab:delay} shows the percentiles of delays over the population of vehicles. For example, 99.5\% of all vehicles are delayed four or fewer ticks.

We adopt percentile values as a metric for fairness because the distribution of latency has a long tail. Specifically, agents who suffer excessively would be those concentrated in the high percentiles even though the mean delay may not vary much between fair and unfair outcomes. 
\revised{We compute \emph{skewness} ($\gamma$) and \emph{kurtosis} ($\kappa$) for the delay distribution. }
Skewness is defined as a measure of symmetry or lack of it \cite{Croarkin+12:NIST-handbook}. A distribution is symmetric if it looks the same to the left and right of the center point. A distribution with a lower (closer to zero) skewness value is fairer. 
Kurtosis is a measure of whether the data are heavy-tailed or light-tailed relative to a normal distribution  \cite{Croarkin+12:NIST-handbook}. 
A heavy-tailed distribution for the time delay indicates that a small number of agents suffer excessive time delay. A distribution with low kurtosis (light-tailed) for the time delay is fairer than a distribution with high kurtosis (heavy-tailed). 

\begin{figure}[!htb]
\centering
\begin{tikzpicture}
\begin{axis}[
    height=6cm, 
    width=8cm,
    ybar,
    bar width = 0.7cm,
    ymin=0,
    xmin=3.5,
    xmax=6.5,
    ymin=0,
    ymax=2800,
    ytick={0, 600, 1200, 1800, 2400},
    y grid style={densely dotted, line cap=round},
    ylabel=\# of vehicles,
    xtick={4,5,6},
    xlabel=Delay (in number of ticks),
    nodes near coords,
    every node near coord/.append style={font=\small},
    nodes near coords align={vertical},
]
\addplot 
[
fill=blue!80!black!60,
postaction={pattern=grid},
] 
coordinates {(4,2117) (5,973) (6,425)};
\addplot[fill=red!80!black!60,
]
coordinates {(4,1917) (5,948) (6,650)};
\legend{\framework,\frameworkNeu}
\end{axis}
\end{tikzpicture}
\caption{
Benefit from prosociality. Delays ($\geq$ 4) with \frameworkNeu and \framework.
35\% of vehicles that experience the highest delay (6 ticks) with \frameworkNeu experience an improvement of $\geq$ 16.67\% with \framework.
}
\label{fig:prosociality-delays}
\end{figure}
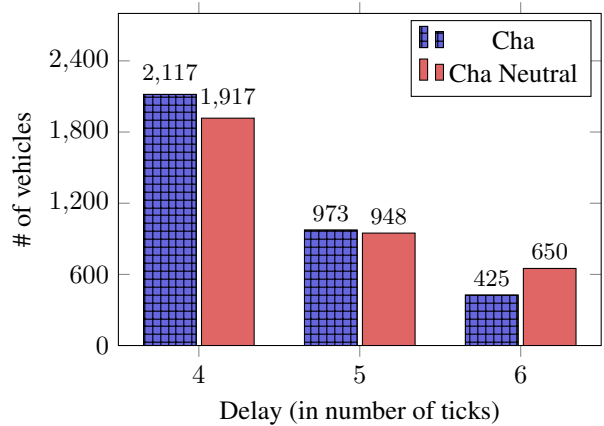

\revised{
Figure~\ref{fig:prosociality-delays} shows the delays of four or more ticks with and without prosociality.
We observe that, by reduction, \framework which incorporates prosociality, improves the outcome for 35\% of vehicles that experienced a worst-case delay of six ticks. 
Specifically, 225 (that is, 35\%) of the 650 vehicles which experienced the highest delay (six ticks) under \frameworkNeu (without prosociality), experience a delay of five ticks or fewer under \framework \emph{with prosociality}. Each of these vehicles saw a delay reduction of at least $\frac{6-5}{6}$ = 16.67\%. \framework thus promotes Rawls' \shortcite{Rawls-99:justice} Maximin doctrine.}
\revised{Also, note that, considering the values in Figure~\ref{fig:prosociality-delays}, we have 425 ticks less delay for \framework \emph{with prosociality} (i.e., $[2117\times4 + 973\times5 + 425\times6] - [1917\times4 + 948\times5 + 650\times6] =-425$). So not only \framework \emph{with prosociality} improves the total mean delay, but also it impacts the heavy tail.}

\begin{table}[!htb]
\caption{Delays with \frameworkNeu and \framework.}
\label{tab:delay}
\centering
\begin{tabular}{l rrrrr rr}
\toprule

& \multicolumn{5}{c}{\emph{Percentiles}} & \multirow{2}{*}{$\gamma$} & \multirow{2}{*}{$\kappa$} \\ \cmidrule{2-6}
& {99}& {99.5}& {99.7}& {99.9}& {100} & & \\ \midrule

\frameworkNeu &3 &4 &5 &6 &6 & 2.07 & 4.99 \\ 
\framework  &3 &4 &\textbf{4} &\textbf{5} &6 & 2.02 & 4.55 \\\bottomrule
\multicolumn{8}{p{2in}}{$\gamma$ = Skewness; $\kappa$ = Kurtosis}

\end{tabular}
\end{table}

\revised{We also compute the Gini coefficient (\fsl{Gini}) to measure the degree of disparity in a distribution. A lower \fsl{Gini} indicates a lower degree of disparity. Table~\ref{tab:disparity-gini} lists \fsl{Gini} for the vehicles in 99, 99.5, 99.7, and 99.9 percentile population. We observe that \fsl{Gini} for 99, 99.5, and 99.7 percentile populations is lower for \framework compared to \frameworkNeu, indicating lower disparity. However, \fsl{Gini} for 99.9 percentile population of vehicles is higher in \framework compared to \frameworkNeu. Under \frameworkNeu, all vehicles face the worst-case delay (six ticks), and thus the disparity is lower, whereas under \framework, the delay is either five ticks or six ticks.}

\begin{table}[!htb]
\caption{\revised{Disparity with \frameworkNeu and \framework. Lower is better.}}
\label{tab:disparity-gini}
\centering
\begin{tabular}{@{}l SSSS@{}}
\toprule
\diagbox{Gini}{Percentiles}  & {99} &  {99.5} & {99.7} &  {99.9} \\\midrule
\frameworkNeu & 0.1896277508046399 & 0.20160427135584336 & 0.15705822906522113 & 5.867613055776399e-17 \\
\framework & 0.16653354651638128 & 0.17098612188429216 & 0.148150077228366 & 0.10910542249352645 \\
\bottomrule
\end{tabular}
\end{table}

\subsection{Testing the Social Welfare Hypothesis}
\label{sec:subresult3}

\paragraph{Hypothesis.}
We take Hypothesis H\fsub{social} as stating that \framework wins (in aggregate societal
gains) over representative central and hybrid approaches. \revised{The corresponding null hypothesis indicates no gains under \framework.}

\paragraph{Metric.}
Our metric is mean travel time for the whole intersection. It is calculated by adding the best-case travel time (that is, 19 ticks, one for each cell in each lane) to the mean delay (mean number of stops for all vehicles in each tick.

\paragraph{Baselines.} 
There are two main traffic control paradigms, pretimed (in which the signals change in a fixed pattern) and actuated (in which signals change in response to vehicles arriving at the intersection from different directions). The actuated paradigm is known to deliver higher performance than pretimed. Therefore, we adopt a variant of the actuated paradigm, called the fully actuated control strategy \cite{Urbanik+15:traffic-signals} as one of our baseline approaches. To compare to previous multiagent systems research, we also adopt the Silk \cite{IJCAI-16:Silk} approach as another baseline because Silk is a hybrid approach and studies a similar social dilemma situation in a similar traffic setting. \revised{We highlight the difference between \framework and Silk in  Section~\ref{sec:related}.} 

Fully actuated control is a traditional approach to regulating traffic. In  fully actuated control, each \emph{movement}, for example, the northbound traffic flow at an intersection, has a detector. A \emph{phase} is a combination of nonconflicting movements. We pair northbound and southbound flows to create one phase and eastbound and westbound flows to create the other. We assume  fully actuated control alternates between these phases. Each phase has a \emph{minimum green} (set to one in our experiment); thereafter, the green can be extended indefinitely provided a vehicle keeps on being detected at the intersection.

\revisedst{Silk is a hybrid framework in which agents learn norms but with hard integrity constraints (called ``laws'') imposed on them. A law can prevent certain actions by the agent. Silk includes a central generator that recommends norms to the agents: a norm emerges if the agents accept the recommendation. Silk supports only static settings.}
For our comparison \revised{with Silk's framework}, we used the payoffs of Table~\ref{tab:payoff-matrix} in both Silk and \framework. 
Since Silk supports only static settings with fixed payoffs, we set the payoff of \emph{Stop} in Silk to zero.

Elaborating on Hypothesis H\fsub{efficient} of Section~\ref{sec:efficient}, we consider a specific hypothesis that \framework yields a lower mean travel time than both Silk and fully actuated control. These experiments are amenable to statistical hypothesis testing: the null hypothesis is that there is no significant difference in mean travel times yielded by \framework, Silk, and fully actuated control. We run these experiments for the case where the north-south orientation has 30\% higher traffic than the east-west orientation.

\begin{figure}[!htb]
    \centering
\begin{tikzpicture}
  \begin{axis}
    [
    title = {},
    xlabel = {Average delay},
    boxplot/draw direction=x,
    height=6cm, 
    width=8cm,
    ytick={1,2,3},
    yticklabels={\framework, Silk, Fully actuated control},
    y tick label style={align=center}
    ]
    \addplot+[
    boxplot prepared={
      lower whisker=0.47,
      lower quartile=0.71,
      median=0.975,
      average=1.03,
      upper quartile=1.337,
      upper whisker=1.56
    }, 
    ] coordinates{};
    \addplot+[
    boxplot prepared={
      lower whisker=2.19,
      lower quartile=3.03,
      median=3.72,
      average=3.65,
      upper quartile=4.24,
      upper whisker=4.78
    }, 
    ] coordinates{};
    \addplot+[
    boxplot prepared={
      lower whisker=3.08,
      lower quartile=4.44,
      median=5.45,
      average=5.53,
      upper quartile=6.45,
      upper whisker=8.07
    }, 
    ] coordinates{};
    \end{axis}
\end{tikzpicture}
    \caption{Comparing average delay in \framework ($\mu=0.98$), Silk ($\mu=3.65$), and fully actuated control ($\mu=5.53$). The benefits of the prosocial approach are clear.}
    \label{fig:boxplot-delay}
\end{figure}
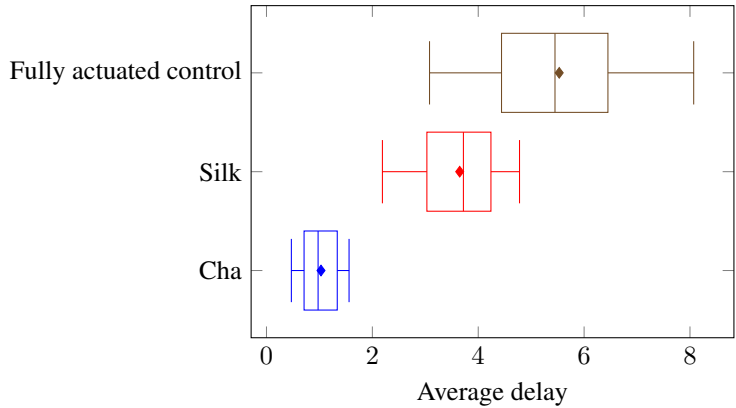

Figure~\ref{fig:boxplot-delay} compares the average delay in \framework, Silk, and fully actuated control.
Table~\ref{tab:travel-time} summarizes the travel time results.
It lists the best-case travel time, mean travel time ($\mu_{travel}$), and standard deviation ($\sigma_{travel}$) for \framework, Silk, and fully actuated control. Table~\ref{tab:travel-time} also lists the $p$-value from two-tailed $t$-test assuming unequal variances, effect size via Glass' $\Delta$ \cite{grissom2012effect}, which is measured as the difference in means divided by the standard deviation of the control group), and \% improvement obtained by \framework over Silk and fully actuated control. We choose Glass's $\Delta$ to measure effect size since the standard deviations ($\sigma$) for the two groups are different.

We find that \framework reduces the delay and yields significantly less travel time than both Silk ($p<0.01$; $\Delta=2.06$) and fully actuated control ($p<0.01$; $\Delta=1.76$), reducing it by 11.79\%  over Silk and 18.55\% over fully actuated control.
Following Cohen's \shortcite{Cohen-88:Statistics} guidelines, an effect of over 0.80 is large; the above effect sizes are substantially above Cohen's guideline.

\begin{table}[!htb]
\caption{Travel time: \framework versus Silk and Fully actuated control.}
\label{tab:travel-time}
    \centering
    \begin{tabular}{lrrr}
        \toprule
        & \framework  & Silk & Fully actuated control \\\midrule
        Best case & 19.00  & 19.00 & 19.00 \\
        Mean $\mu_{travel}$ & 19.98 & 22.65 & 24.53 \\
        Stdev $\sigma_{travel}$  & 0.55 & 1.28 & 2.56 \\\midrule
        \% Improvement &--&11.79 & 18.55 \\
        $p$-value &--& $<0.01$ & $<0.01$ \\
        Glass' $\Delta$ &--& 2.06  & 1.76  \\
        \bottomrule
    \end{tabular}
\end{table}


\section{Discussion}
\label{sec:concl}

\framework is a flexible and dynamic framework for norm emergence in multiagent systems that supports prosocial outcomes. In \framework, each agent reasons individually about which norms to develop. Our results support our hypotheses.

One, norms that emerge in \framework resolve conflicts efficiently  and improve system-level outcomes.
Two, \framework is responsive to changes in the environment.
%
%
Three, more importantly, in light of recent increasing awareness to promote and foster prosociality in artificial intelligence generally and multiagent systems, \framework supports prosocial outcomes, specifically, fairness in resource allocation \cite{AAMAS-20:Blue-Sky}. \framework incorporates prosociality based on inequity aversion and naturally respects Rawls' Maximin doctrine to improve the worst-case outcome on members of the society.
To our knowledge, no other approach tackles the concerns of prosociality in the study of norm emergence \revised{in multiagent systems}.
Four, \framework yields higher societal gains than both a representative central approach and a hybrid approach (baselines in our evaluation).

Note that \framework applies reinforcement learning to norms. Unlike work on social
learning, \framework shows how incorporating guilt disutility can lead
to prosocial decision making and thus to fair societal outcomes.
Also, \framework does not require repeated interactions in a fixed graph. 
Departing agents can convey their experience to those who come
after them, which facilitates norm emergence by giving endurance to
the system even though individual agents have short lifetimes. 
Note that \framework automatically garbage collects useless
norms. When the system changes, new norms emerge, and some old
norms are unlearned (they lose utility and are no longer selected).

It is worth mentioning that \framework being a
dynamic and decentralized framework, not only supports beginning
from an initially empty set of norms but can also be used when considering an initial set of existing norms, on top of which the agents can then build new norms.

\revised{
A potential critique of this work is that even though the agents are decentralized, there is a system-wide notion of conflict that leads to their behaviors being coordinated. Arguably, this setup accords with \citepos{Elsenbroich+Verhagen-16:agents} idea of how functionality and complexity may shift between an agent and its context.
However, the conflicting situations in our study are motivated by the domain of traffic and earn their legitimacy from that domain. In another domain, the actions and conflicts would be different. But the conflicts do not determine the outcomes and other approaches with a similar conflict structure can produce different results.
}

\subsection{Related Work}
\label{sec:related}

It is worth highlighting how \framework differs from the recent literature on norm emergence. \revised{Note that we do not intend this article to be a survey of norms or normative multiagent systems in general: see, for instance, \citet{Hollander}, \citet{Nardin+16:Sanctioning}, and \citet{Andrighetto2013}.}

\paragraph*{Static and Closed.} Many of the existing game-theoretic approaches for social norm
emergence consider agents in a closed system such as a graph
structure, and fixed payoffs are considered for the agents' behaviors \cite{Airiau2014,sugawara-AAMAS14-emergence,Sen-IJCAI07-NormEmergence,Villatoro-IJCAI11-SocialInstruments,Beheshti+AAAI15-Cognitive}. In these types of studies, agents are nodes of a graph, and the interactions between them create the edges of the graph. Agents learn their policies by playing repeated games with random agents. Through the course of these repeated games, new norms emerge in the system. In contrast, \framework considers a system that is open (agents continually leave and enter), dynamic, and addresses prosociality.

\paragraph*{Static and Centralized.}
Other works consider norm emergence in a setting where a central agent with complete knowledge of the systems generates (or dictates) the norms. 
Morales {\etal} \shortcite{Morales:2015} proposed a mechanism called IRON along with two other variations \cite{Morales-AAMAS14-Minimality,Morales-AAMAS15-LiberalNorms} for the on-line synthesis of norms. IRON, however, could result in unstable normative systems. To mitigate the instability problem, Morales {\etal} \shortcite{Morales2018} presented an offline method called ``System for Evolutionary Norm SynthEsis'' (SENSE), which builds on top of IRON. 
These mechanisms employ a central approach that relies on global knowledge. The central agent observes the interactions of members to synthesize conflict-free norms in a top-down manner. \revised{Even though these frameworks share some commonalities with \framework in terms of norm representation and norm generation (albeit static), they are fundamentally different from it. Whereas these mechanisms use a central norm generator and learner, \framework uses a dynamic and decentralized mechanism for its four phases, as described in Section~\ref{sec:Framework}. Additionally, \framework agents incorporate prosocial decision making to achieve fairness. In this manner, \framework brings together two important themes in prosociality: decision making by individuals and fairness of system-level outcomes.}

\paragraph*{Static and Decentralized}
Hao {\etal} propose two strategies based on local exploration and global exploration to facilitate norm emergence in a bottom-up manner \cite{Hao+TAAS18-efficient}. 
Whereas Hao {\etal}'s strategies focus only on maximizing average payoff for all agents, i.e., social welfare, \framework's focus is on maximizing both social welfare and fairness. 
Mihaylov {\etal} \shortcite{Mihaylov2014} propose a decentralized approach for convention emergence in multiagent systems for a pure coordination game with fixed payoffs. Once they have learned a convention, agents will not re-learn it (i.e., have a static environment). 

\paragraph*{Static and Hybrid}
\revised{Silk \cite{IJCAI-16:Silk} is the closest framework to \framework in the literature. Silk is a hybrid framework for the regulation of open normative systems. Briefly, this framework is composed of a central generator, which imposes hard integrity constraints 
that agents must follow, and recommends norms as soft constraints. Thus, a norm can emerge based on the decision of various agents to accept or deny the recommendations. In addition, Silk's framework supports only static settings with fixed payoffs. Moreover, Silk tackles the problem of fairness through its central generator which continually monitors the performance and intervenes if necessary to improve fairness, whereas \framework agents learn to be prosocial.}

\paragraph*{Dynamic.}
Knobbout {\etal} \shortcite{Knobbout2014} propose update semantics for norm addition to characterize the dynamics of norm addition in a formal way. Updates are parameterized over actions. However, this work does not indicate whether its proposed model can be applied to regulate agent's actions and satisfy system-level goals.
\revised{Verhagen \cite{Verhagen-2001-simulation} puts worth the need for flexible and decentralized system. The simulation in Verhagen \cite{Verhagen-2001-simulation} focuses on spreading and internalization of norms in such systems. However, Verhagen assumes that a top-level entity (say, a normative advisor) is aware of the correct norm, and this group norm (g-norm) does not change. This g-norm is spread to the agents through the normative advice provided using a top-down approach by a centralized authority (i.e., there is a leader in the society). In contrast, \framework does not enforce a need for a central authority and the norms in \framework are not fixed.}
\revised{Savarimuthu {\etal} \shortcite{Savarimuthu2009, Savarimuthu2010ObligationNI} propose an architecture where agents can identify norms of the society in a bottom-up fashion. They address how an agent might be able to identify whether a norm is changing in society and how it might react to this situation. However, \framework is different from this work in that \framework is an on-line framework; that is, in \framework, norms are learned rather than inferred through data mining. }

\paragraph*{Prosocial}
Recently there has been increasing interest in designing multiagent systems that lead to prosocial behaviors in agents \citep{Santos2019EvolutionOC,Paiva2018EngineeringPW,Serramia2018,AAMAS-20:Elessar}.
Prosocial behavior is when an agent performs an action that is unfavorable to itself but benefits others. 
We consider the problem of agents prosocially acting while sharing resources in a multiagent system. The connection between norms and prosociality, though conceptually important, has not received adequate attention: existing AI approaches on prosociality focus on individual decision making by agents. In contrast, we relate social norms both to prosocial decision making and to societal outcomes such as fairness. Specifically, the norms through which agents coordinate their interactions may or may not be fair.

\revised{Table~\ref{table:comparison table} summarizes the difference between \framework and some leading frameworks addressing a similar setting.}

\begin{table}[!htb]
\caption{\revised{Comparison of characteristics of simulation works on norms (Yes -- Considered; No -- Not Considered; NA -- Not Applicable; NS -- Not Specified)}
}
\centering
\resizebox{\columnwidth}{!}{%
\begin{tabular}{l c c c c c c}
\toprule
Model & Creation & Identification & Architecture & Openness &Dynamism & Prosocial\\ [0.5ex] 
\midrule
\framework &on-line&RL&Decentralized & Yes & Yes &Yes\\
Morales {\etal} \cite{Morales:2015, Morales-AAMAS14-Minimality,Morales-AAMAS15-LiberalNorms}&on-line&Case-based Reasoning&Centralized & No & No & No \\
\citet{IJCAI-16:Silk}  &on-line&RL &Hybrid&Yes&No&Yes\fsup{*}  \\
\citet{Sen-IJCAI07-NormEmergence}  &on-line&RL&Decentralized&No&No&No  \\
\citet{Airiau2014}  &on-line&RL&Decentralized&No&No&No  \\
 \citet{Villatoro-IJCAI11-SocialInstruments} &on-line&RL&Decentralized&No&No&No  \\
Savarimuthu {\etal} \shortcite{Savarimuthu2009, Savarimuthu2010ObligationNI} & \revisedb{on-line} & Data Mining & Decentralized & Yes & Yes & No \\
Beheshti {\etal} \shortcite{Beheshti+AAAI15-Cognitive} & on-line & Game Theory \& RL & Decentralized & No & No & No \\
Frantz {\etal} \shortcite{Frantz2013} & on-line & RL & Decentralized & No & Yes & No \\
\citet{Hao+TAAS18-efficient} & on-line & RL & Decentralized & No & No & No \\[1ex]
\bottomrule
\multicolumn{4}{l}{\fsup{*}\footnotesize{Enforced by Central Generator}}
\end{tabular}%
}
\label{table:comparison table}
\end{table}

\subsection{Prosociality in General}


Applications of our approach on practical domains would be important to validate it in broader settings.

\revisedb{Once following certain conditions are met, \framework can be successfully applied to other settings besides the traffic one described here:}
\begin{itemize}
\item \emph{Agents have asymmetric interests.} \framework assumes its population of users has asymmetric interests that may lead to conflicts.

\item \emph{Conflicts must be detectable}. Norm generation in \framework is based on detection of the conflicts (e.g., car collision), so \framework assumes conflicts are detectable.

\item \emph{Some agents’ preferences can be assumed at the design time}. \framework assumes some domain information such as payoffs are provided by the system designer (e.g., the payoff of car collision).

\item \emph{Agents can observe and communicate}. \framework assumes its members are able to share their experience to agents of the same type. Additionally, agents of different types are able to receive and send accumulated costs in order to learn and act prosocially.

\end{itemize}

\revisedb{
A natural family of application scenarios for \framework is the development of what \citet{Berners-Lee-99:Weaving} calls \emph{social machines}. We can think of social machines as sociotechnical systems  \cite{WWW-16:IOSE} geared to support interactions between people or people and businesses in reference to some shared resources or coordinated decision making. Social machines today are largely not formulated to satisfy our desiderata but they could and should be. Today, there is little computational support for prosociality, flexibility, or dynamism and the underlying architecture being based on servers is decidedly centralized. Thus, these desiderata are left to the human participants to achieve informally. But we can envision a situation where decentralized agents assisted the humans and provided computational support for prosociality, flexibility, and dynamism.
}


\revisedb{Another domain of application for \framework is on-line communities, in which members continually interact by exchanging information (e.g., comments and
pictures) in different forums. 
An on-line community is an open, dynamic system. It is open because the users can enter the system T any time. It is dynamic because users may also change their preferences and behaviors. 
The goal is to achieve ``healthy''
communities \shortcite{Hinds2011}.  Specifically, the goal is
to avoid circumstances in which a user expresses concern about the content involving that user that is shared by other users. Such content may include pictures that violate one user's privacy but may be funny in the eyes of the posting user \cite{TOCHI-17:Multiuser}.
Prosociality would be demonstrated by the posting user taking into account the preferences of the people shown in a picture.
Flexibility is indicated by the users not having rigid stances for their preferences, which can shift depending on the context. 
Endurance is indicated by the users interacting multiple times, sometimes over the course of years.
Dynamism is indicated by the fact that norms for what is acceptable (e.g., in regards to sensitive information) may change depending on the cultural or political situation.
Decentralization is indicated by the users deciding individually on norms without any computational limitations imposed by any sharing platform they may be using.
The users could concede to each other and positively or negatively sanction each other. }

The domain of Open Source Software Development \revisedb{(OSSD)} communities
\cite{Avery+16:mining} is particularly compelling because it brings
together considerations of ethics along with inertia (projects can
last years) and group norms. 
\revisedb{An OSSD community is open and dynamic. With a defined goal, one or more developers start the development of an open source software project. Other developers who are interested in the project, contribute to the project's goal by writing software code. 
Prosociality in OSSD would be demonstrated by the community of developers (and users outside of the developer community) identifying issues with the ongoing project and voluntarily contributing fixes to resolve the identified issues. 
Similar to in on-line communities, endurance in OSSD communities is indicated by the users interacting multiple times over the course of years. 
Flexibility is indicated by the developer community allowing users to submit feature requests and subsequently redefining the project's goals. 
Dynamism is indicated by the fact that individual and group norms as well as the project's goals may evolve over time as the community grows.
Decentralization is indicated by the OSSD community developers individually deciding on the norms on how they interact with other developers and users, and how they sanction them.
}

\revisedb{
Another example scenario where the ideas of \framework could apply is in shared transportation. \citet{IC-Ethics-20:microtransit} describe the setting of public microtransit (i.e., last-mile ride sharing) in which a public van service is used to help members of the public who cannot use a private car (e.g., because of its expense or their poor health). This setting demonstrates the features necessary to achieve the desiderata identified for \framework.
Prosociality would be demonstrated by riders adjusting their requested ride's origin, destination, or timing to accommodate the needs of other riders.
Flexibility is indicated by the riders changing how they interact, e.g., in response to environmental events (is it raining) or their health (does one of them need to get to a dialysis appointment?). 
Endurance is indicated by the riders interacting with each other again and again (e.g., every weekday), sometimes over the course of years, because the people who need such assistance in a neighborhood are likely to be sharing rides more than once.
Dynamism is indicated by the fact that norms for what is acceptable may change depending on the public health (e.g., in regards to an epidemic flaring up or dying down) or economic (e.g., during a recession, money is tight and schedules may be more rigid for workers) situation.
Decentralization is indicated by the riders deciding individually on norms on how they interact.
The users could help each other, thank each other for accommodating each other's requests, or negatively sanction each other, e.g., through shunning each other's company.
}

\subsection{Outlook and Directions}
An important future direction is to understand the context as a basis for overriding norms. For instance, can we allow vehicles with a sick passenger to Go when the norm suggests Stop?  Such overrides may be facilitated by sharing context and explanations to justify the ethicality of deviating from a norm \cite{IJCAI-18:Poros}.

Another direction is to include more complex norm actions, for example, (1) communicating an obligation to Stop to let an ambulance pass; and (2) if in a rush, delegate (to the vehicle behind) a norm to help a stranded vehicle. One way to model more complex norm actions is through sanctioning, which also corresponds to an additional (state, action) pair in the payoff matrix.

Understanding societal inertia in the sense of how new norms can supersede existing norms is important; \framework is dynamic but how to evaluate and how to reduce the societal friction and inertia in arriving at new norms remains to be studied. 

\framework has a flavor of group norms since agents of the same type pass on their experiences with other agents of the same type. But each agent individually reasons whether to follow an applicable norm or not. Another natural line of future research is to explore the generation of group norms as opposed to individual norms \cite{Alechina+17:decomposition,Aldewereld-TAAS16-GroupNorms} bring together notions of group intentions \cite{Dunin-Keplicz+Verbrugge-10:teamwork,Ecai-98} and group ability \cite{Self90:maamaw-book}. Group norms apply to groups of agents together. 
For example, in the world of autonomous cars, car agents sharing the same space can share information with each other and can decide actions and move together as a platoon of cars (i.e., creating a group norm). Group norms raise important challenges of how a group can allocate decision-making authority and accountability and whether the actions taken on behalf of a group satisfy ethical criteria both with respect to group members and with respect to outsiders.



\section*{Acknowledgements}
We would like to thank Emily Berglund, Rahmatollah Beheshti, and Weijia Li for helpful discussions. We would also like to thank the anonymous reviewers for their careful reading of our manuscript and their insightful comments and suggestions.
NA acknowledges support by the Department of Defense through the Science of Security Lablet at North Carolina State University. MPS acknowledges partial support from the NSF under grant IIS-2116751. 


\bibliographystyle{plainnat}
\bibliography{Munindar,trafficj,Nirav}

\end{document}